\documentclass[a4paper,12pt,epsfig]{article}
\usepackage[dvips]{graphicx}
\usepackage{amstext}
\title{\bf
Evidence for an Exotic Baryon State, $\Theta^+(1540)$, in
Photoproduction Reactions from  Protons and Deuterons with CLAS}

\author{Valery~Kubarovsky$^{*\dag}$ and Stepan~Stepanyan$^{\dag}$}

\begin{document}
\date{}
\maketitle

\begin{center}

\vspace{-0.5cm}
{\em $^{*}$Rensselaer Polytechnic Institute, Troy, NY  12180-3590,\\
$^{\dag}$Jefferson Lab, 12000 Jefferson Ave., Newport News, VA 23606\\[5mm]}
{\large and the CLAS Collaboration}\\[5mm]
Talk at CIPANP 2003, New York, May 19-24,2003
\end{center}

\begin{abstract}
CLAS photoproduction data on deuterium and hydrogen targets have
been analyzed in a search for an exotic baryon state
with strangeness $S=+1$,  
the $\Theta^+$ (originally named the $Z^+$). 
This resonance was predicted recently in  theoretical work based on
the chiral soliton model as a lowest mass member of an anti-decuplet of
5-quark states. The reaction $\gamma d\to pK^-K^+n$,
which requires a final state interaction inside the deuteron, was used
in the analysis of deuteron data. In the
analysis of proton data, the reaction 
$\gamma p \rightarrow \pi^+K^-K^+n$
was studied. Evidence for the $\Theta^+$ state is 
found in both analyses in the invariant mass distribution of the
$nK^+$. 
Our results are consistent with previously reported
results by LEPS/Spring-8 collaboration (Japan), and by
the ITEP (Moscow) group.
\end{abstract}

\maketitle

\section{Introduction}
\indent

Pentaquark resonances have been predicted
decades ago  and there have been experimental searches for many years.
However, no significant signal was found in the early work.
Recent theoretical work based on the chiral soliton model
\cite{DIAKPOL} made more quantitative predictions for the masses and
widths of a spin ${\bf s}=1/2$ anti-decuplet of 5-quark states 
($qqqq\bar q$). Using the $P_{11}(1710)$ resonance as
the ``anchor'' for the masses
of the anti-decuplet, the lowest lying member, $\Theta^+$, is
predicted to have a mass 1530 MeV$/c^2$ and a width of $\sim 10$ MeV$/c^2$.
 It
is predicted to be an exotic baryon state
with strangeness $S=+1$, and $I=0$. 

The LEPS collaboration at the SPring-8 facility in Japan recently 
reported \cite{nakano} the observation of an $S=+1$ baryon at 
1.54 GeV$/c^2$ with a  width of $<25$ MeV$/c^2$ from the  
reaction $\gamma n \to K^-K^+n$ where the target neutron is bound in 
carbon,  and
the residual nucleus is assumed to be a spectator.
This measurement reported a statistical 
significance of $4.6 \pm 1.0\ \sigma$.  Also,
the DIANA collaboration at ITEP \cite{dolgolenko} recently 
announced results from an analysis of bubble-chamber data 
for the reaction $K^+ n \rightarrow K^0 p$, where the neutron 
is bound in a xenon nucleus, which shows a narrow peak at 
$1539 \pm 2$ MeV$/c^2$.
The statistical significance of the ITEP result is 4.4 $\sigma$. 

The data presented here were taken at the Thomas Jefferson National 
Accelerator Facility with the CLAS detector \cite{clas}
and the photon tagging system \cite{tagger} in Hall B. Data from two
experiments have been used in these analyses: i) photoproduction on
deuterium using tagged photons produced by $2.474$ and $3.115$ GeV
electrons; and ii) photoproduction on protons using tagged photons
produced by $4.1$ and $5.5$ GeV electrons. The exclusive reaction $\gamma d\to
pK^-K^+(n)$ was studied in the analysis of the deuteron data. A peak in the
invariant mass distribution of $nK^+$ was found at $1.542$ GeV$/c^2$ with
a width of $21$ MeV$/c^2$ and the statistical significance 
$5.3 \pm 0.5~\sigma$. The reaction $\gamma p\to \pi^+K^-K^+(n)$ was
studied using the hydrogen data. A peak at $1.537$ GeV$/c^2$ 
with a width of $31$
MeV$/c^2$ in the invariant mass distribution of $nK^+$ was found in this
reaction as well. The statistical
significance of this peak is $4.8 \pm 0.4\sigma$.

\section{Photoproduction on deuterium}
\indent

\begin{figure}[ht]
\begin{center}
\resizebox{28pc}{!}
{\includegraphics[trim=0 85 0 0]{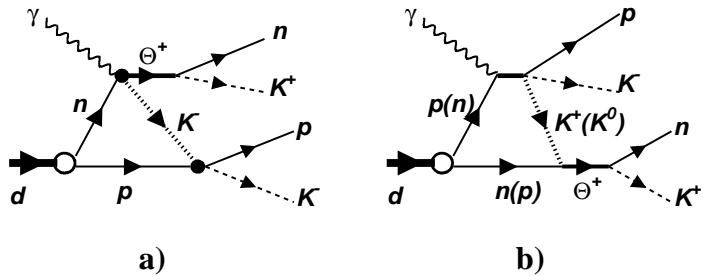}}
\caption{Rescattering diagrams that can contribute to the production
of the $\Theta+$ in the exclusive reaction.}
\label{fig:gdz}
\end{center} 
\end{figure}

In the photoproduction on deuterium the $\Theta^+$ can be produced
directly on the neutron in the reaction $\gamma n\to \Theta^+
K^-$, similar to the reaction mechanism used by the LEPS
collaboration. While the proton is a spectator in the
direct production reaction and will
not be detected in most cases, there are other 
ways to excite
the $\Theta^+$.
Due to the final state interactions 
the proton can obtain high momentum
and be detected. Fig.\ref{fig:gdz} 
shows rescattering diagrams that may
contribute to the production of the $\Theta^+$ in the photoproduction
on deuterium. For identification of such reactions the $p$,
$K^+$, and $K^-$ are detected, and the neutron is identified in the
missing mass analysis. Although these reactions have a smaller cross
section 
compared to the direct production because of an additional rescattering, they
have the following advantages: i) the $K^-$s that are produced predominantly in
the forward direction in the direct production mechanism will
scatter at larger angles and will have higher probability of
detection in CLAS; ii) the kinematics of such exclusive reactions 
puts additional constraints on the event selection that
help to clean up the event sample significantly, and iii) due to the 
exclusive kinematics no Fermi momentum corrections are needed for
the correct calculation of $M(nK^+)$.

\begin{figure}[ht]
\begin{center}
\resizebox{20pc}{!}
{\includegraphics[height=0.2\textheight]{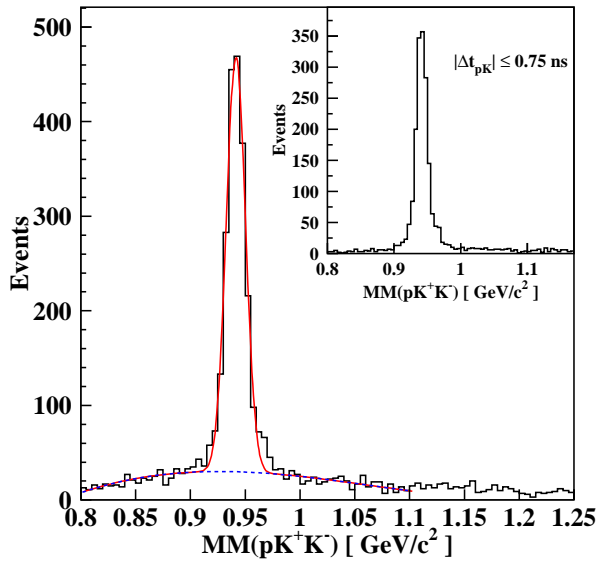}}
\caption{The $MM$ distribution of events after PID cuts. Inset 
shows  $MM$ distribution for events with tight timing cuts for PID.}
\label{fig:mxn}
\end{center}
\end{figure}

The analysis focused on events with one detected proton, 
$K^+$ and  $K^-$ (and no other charged particles) in the final state.
Either the $K^+$ or the $K^-$ in the event was required to have a
time at the interaction vertex within 1.5 nsec of the proton's vertex time. 
Also, the incident photon time at the interaction vertex was required
to be within 
1.0 nsec of the proton to eliminate accidental coincidences.
The missing mass ($MM$) distribution of selected events with 
the $pK^+K^-$ final
state is shown in Fig.\ref{fig:mxn}, where a clear peak at the
neutron mass is seen. 
For further analysis events within $\pm 3 \sigma$ of the
neutron peak were kept. 
Background contributions due to particle misidentification are estimated
at about $\sim 15\%$.
The inset in  Fig. \ref{fig:mxn}
shows the $MM$ distribution of events with a tighter cut on the vertex time
for kaons. Both kaons were required to be within $0.75$~nsec of
the proton vertex time. Practically no background 
remains under the neutron peak in
this case.


There are several known reactions, such as photoproduction of
mesons (that decay into $K\bar{K}$) or excited hyperons (that 
decay into a $pK^-$ or $nK^-$), that contribute to the same final state.  
The $\phi$ meson at $M(K^+K^-)=1.02$ GeV$/c^2$, and the $\Lambda$(1520) at
$M(pK^-)=1.518$ GeV$/c^2$ are cleanly seen in our event sample. 
Events from these 
resonances have been removed from the final sample.

Two other event selection requirements are applied, based 
on kinematics.  
First, the missing momentum of the 
undetected neutron must be greater than 80 MeV/c.  
For momenta below this 
value, the neutron is likely a spectator to other reaction 
mechanisms.  Our studies show that increasing the 
value of this cutoff does not change the final results -- in 
particular it does not eliminate the peak shown below -- but
does reduce the statistics in the $M(nK^+)$ spectrum.
Second, events with $K^+$ momentum greater than 1.0~GeV/c were removed.
This cutoff is based on
Monte Carlo simulations of the 
$\Theta^+$ decay from an event distribution uniform in phase space, which 
show that the $K^+$ momentum rarely exceeds 1.0 GeV/c.  The 
data also show that $K^+$ momenta greater than 1.0 GeV/c are 
associated with an invariant mass of the $nK^+$ system, 
$M(nK^+)$, above $1.7$ GeV/c$^2$.  The second requirement reduces this
background.

\begin{figure}
\begin{center}
  \resizebox{17pc}{!}{\includegraphics{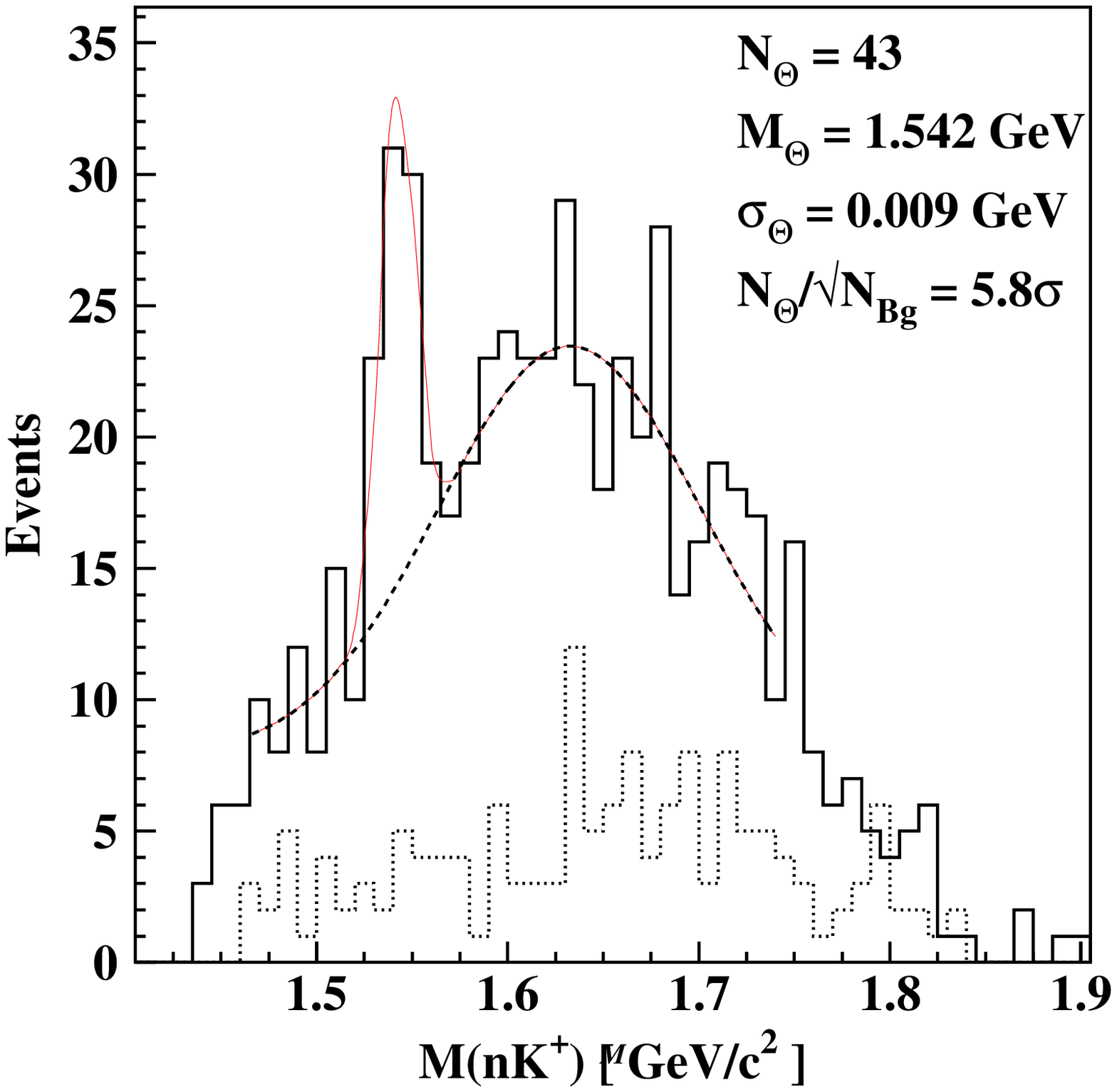}}
  \resizebox{17pc}{!}{\includegraphics{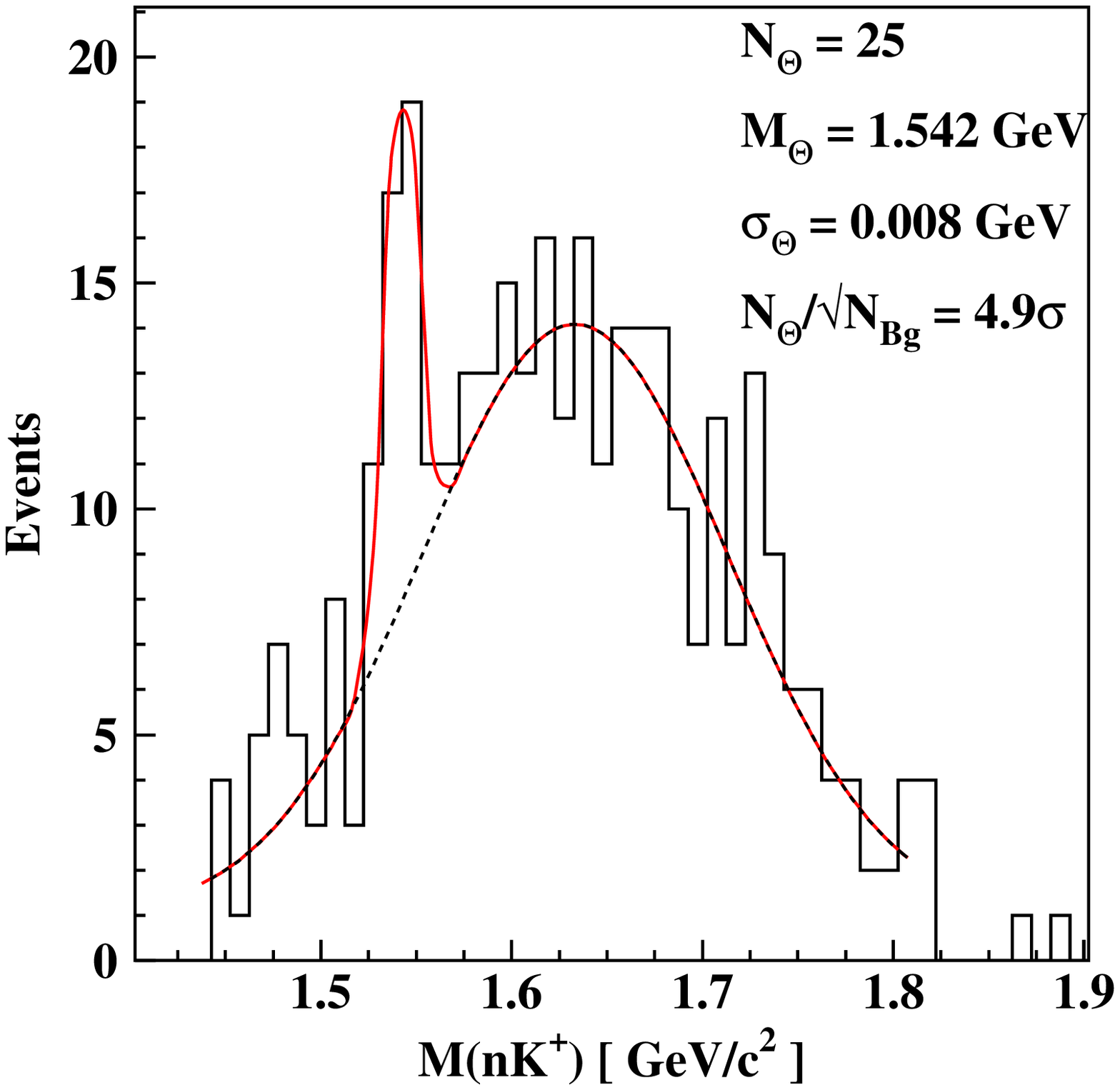}}
\caption{Invariant mass of the $nK^+$ system, which has strangeness 
$S=+1$, showing a sharp peak at the mass of 1.542 GeV/c$^2$.  
The panel on the left corresponds to all selected events, panel on the
right corresponds to events with tight kaon timing cut. The dotted
histogram shows the spectrum of events 
associated with $\Lambda$(1520) production. 
}
\label{fig:final}
\end{center}
\end{figure}

The final $M(nK^+)$ spectrum 
is shown in Fig. \ref{fig:final}, along with a fit to the 
peak at 1.542 GeV$/c^2$ and a Gaussian plus constant term fit 
to the background. The panel on the left shows the distribution for all
selected events. The spectrum of events removed by the
$\Lambda$(1520) cut is shown in the dotted histogram, and  
does not appear to 
be associated with the peak at 1.542 GeV$/c^2$. 
The fit gives a Gaussian width for
the peak 
consistent with the instrumental resolution 
of 21  MeV$/c^2$ (FWHM).
The statistical significance of 
this peak is estimated to be $5.8 \sigma$, based on fluctuations 
of the background over a window of 36 MeV/c$^2$ centered 
on the peak.  
Different assumptions for the background shape lead to an 
additional uncertainty in the statistical significance, 
which is estimated at $5.3 \pm 0.5\ \sigma$.

The panel on the right in Fig.\ref{fig:final} corresponds to events with
a tight timing 
cut on the kaon vertex time. The signal at $1.542$ GeV$/c^2$ is clearly
seen with a smaller number of events, and somewhat reduced
significance. 
This result has since been submitted for publication in 
Phys. Rev. Lett. \cite{clas_theta}.

\section{Photoproduction on the proton}
\indent

\begin{figure}
\begin{center}
\resizebox{30pc}{!}
{\includegraphics[height=1.\textheight]{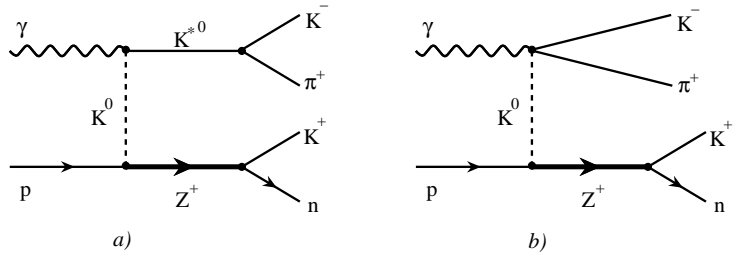}}
\caption{ 
Feynman graphs for the $\Theta^+$ photoproduction from  a proton.}
\label{gamma_p}
\end{center}
\end{figure}

In this analysis the reaction $\gamma p\to \pi^+K^-K^+n$ was
studied. 
Possible diagrams contributing to the
photoproduction of the $\Theta^+$ from  the proton  are
presented in Fig. \ref{gamma_p}. 
An estimate of the cross section for $\Theta^+$ production 
in the reaction $\gamma p\rightarrow K^{*0}\Theta^+$
was made by M.~Polyakov \cite{Maxim}. 
The $d\sigma/dcos\theta_{cm}$ distribution
($\theta_{cm}$ 
is the angle between the $\pi^+K^-$  momentum  
and photon beam in the center of mass system)
peaks in the forward direction
(small $t$ region), as expected for the $t$-channel exchange mechanism 
(see Fig.\ref{gamma_p}).
About 80\% of the events lie in the region with $cos\theta_{cm}>0.5$
(for a photon energy at 4~GeV), 
which appears to be a natural cut for the extraction of the $\Theta^+$ 
signal from a proton target.
This important feature of the cross section
can be used for the signal selection and the background reduction
in  photoproduction reactions.

\begin{figure}[h]
\begin{center}
  \resizebox{18pc}{!}{\includegraphics{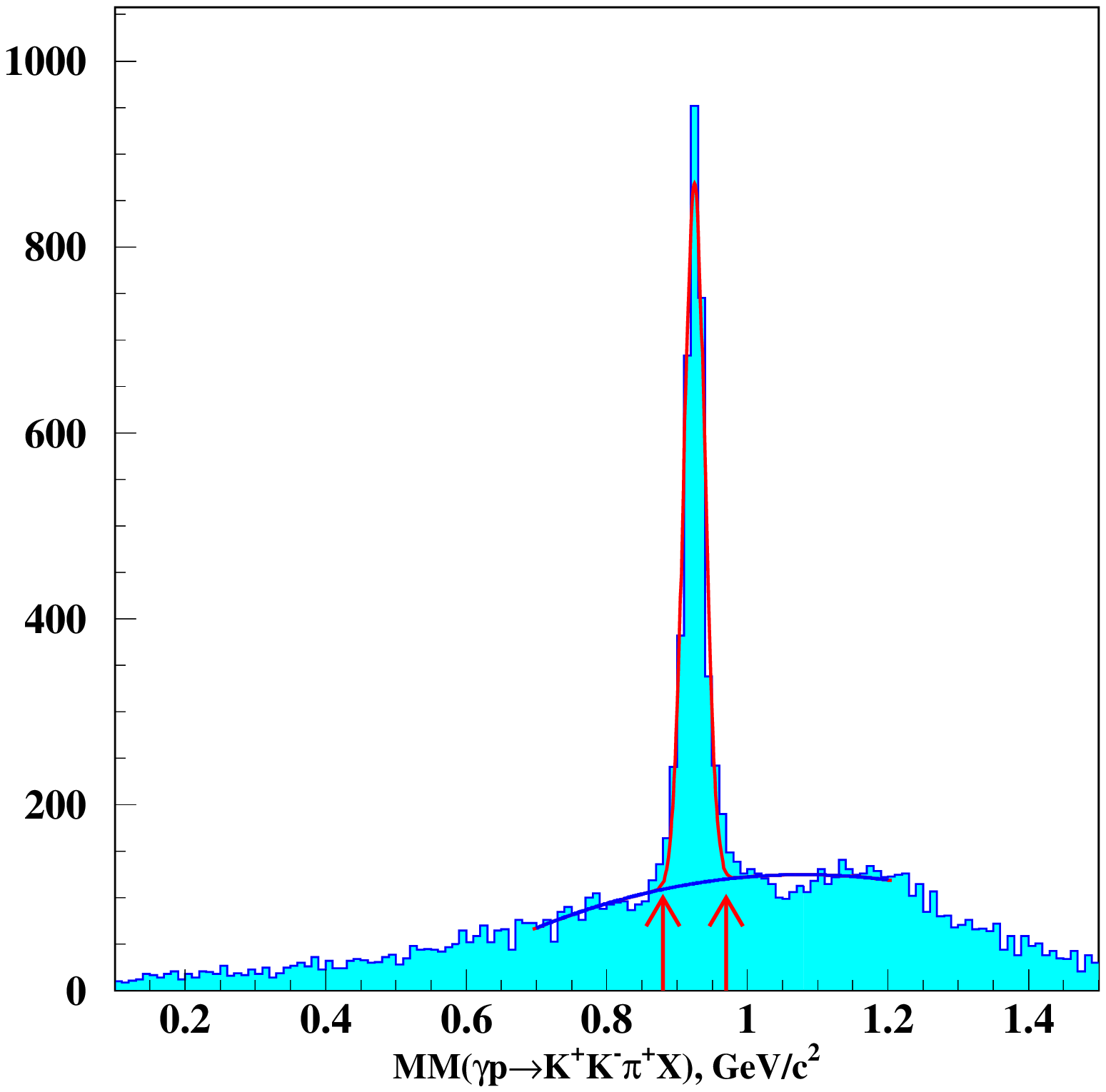}}
  \resizebox{18pc}{!}{\includegraphics{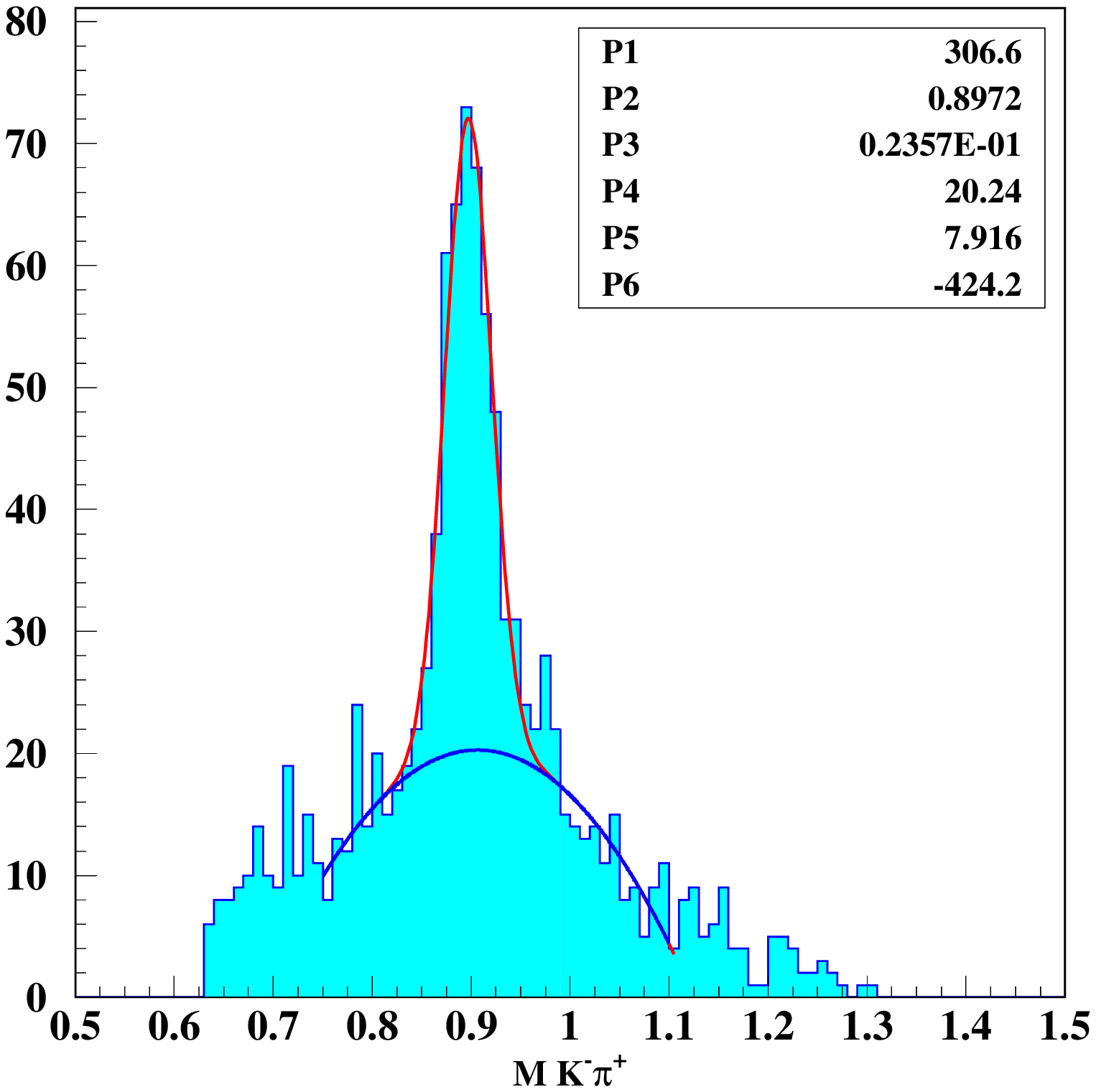}}
\caption{ 
The missing mass 
spectrum for the reaction $\gamma p\rightarrow \pi^+K^-K^+X$.
About 3K neutrons are in the peak.
The arrows indicate the $\pm 3\sigma$ cut for the
the neutron selection.
Right panel: the invariant mass distribution of $\pi^+K^-$ after 
the cut on the neutron
peak.
}
\label{neutron}
\end{center}
\end{figure}

The reaction $\gamma p \rightarrow \pi^+K^-K^+n$ was studied 
at Jefferson Lab with photon energy from 3  to 5.25~GeV 
using an energy tagged photon beam.  
The final state particles,
$\pi^+,~K^-$ and $K^+$, were detected 
in the CLAS detector\cite{clas}, and the neutron was identified 
using the 
missing mass technique.
There are 13.6K events, each  having  a positive pion and 
two kaons of opposite sign 
in the final state, which were selected for the analysis
of the reaction
$\gamma p\rightarrow \pi^+K^-K^+n$.
The missing mass distribution for the reaction 
$\gamma p\rightarrow \pi^+K^-K^+X$
is shown in  Fig.\ref{neutron}.
A neutron  peak is clearly seen in 
this distribution.
There is a 27\% background under the peak. 
The arrows indicate the $\pm 3\sigma$ cut for the neutron selection.

\begin{figure}[h]
  \resizebox{18pc}{!}{\includegraphics{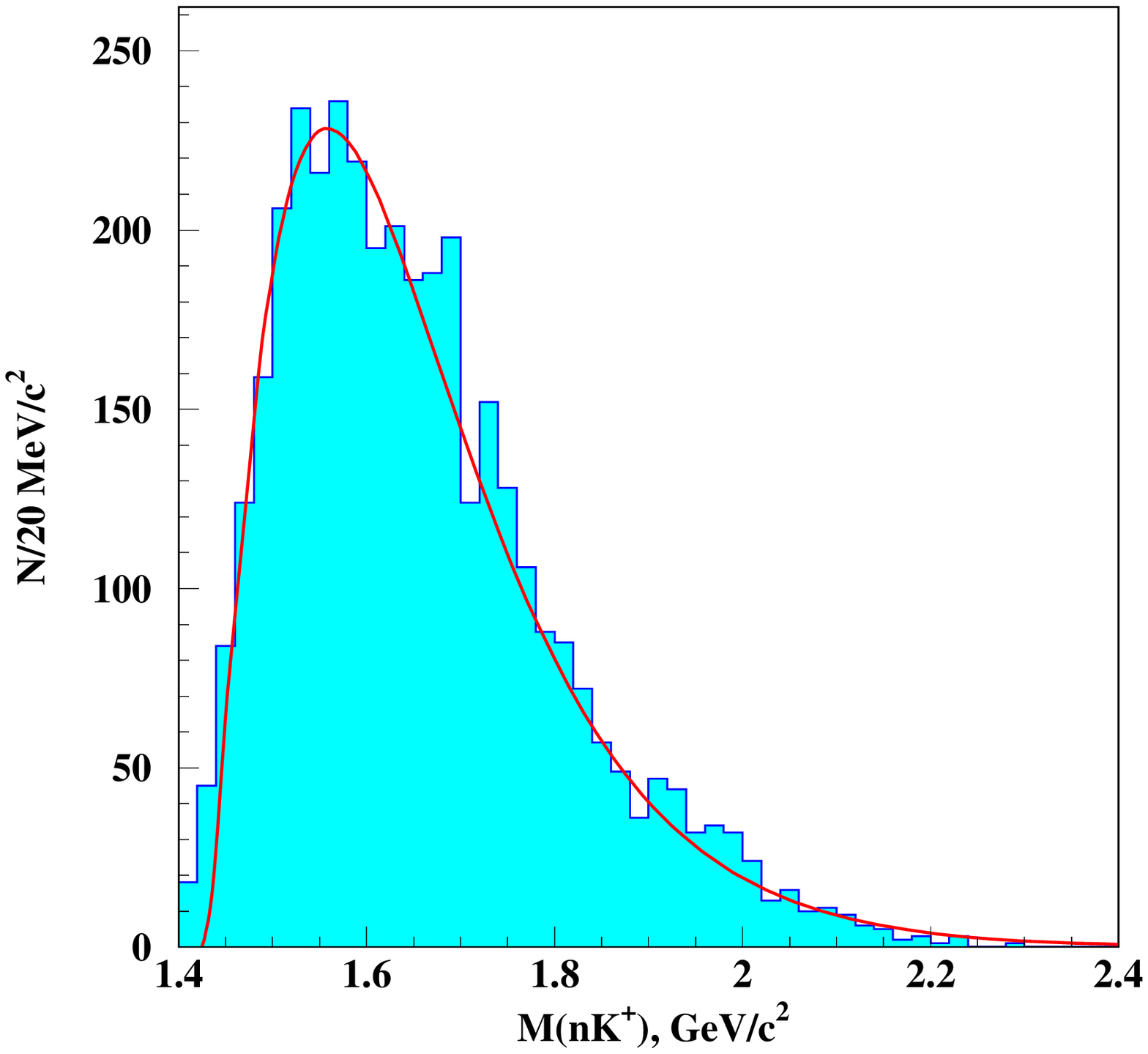}}
  \resizebox{18pc}{!}{\includegraphics{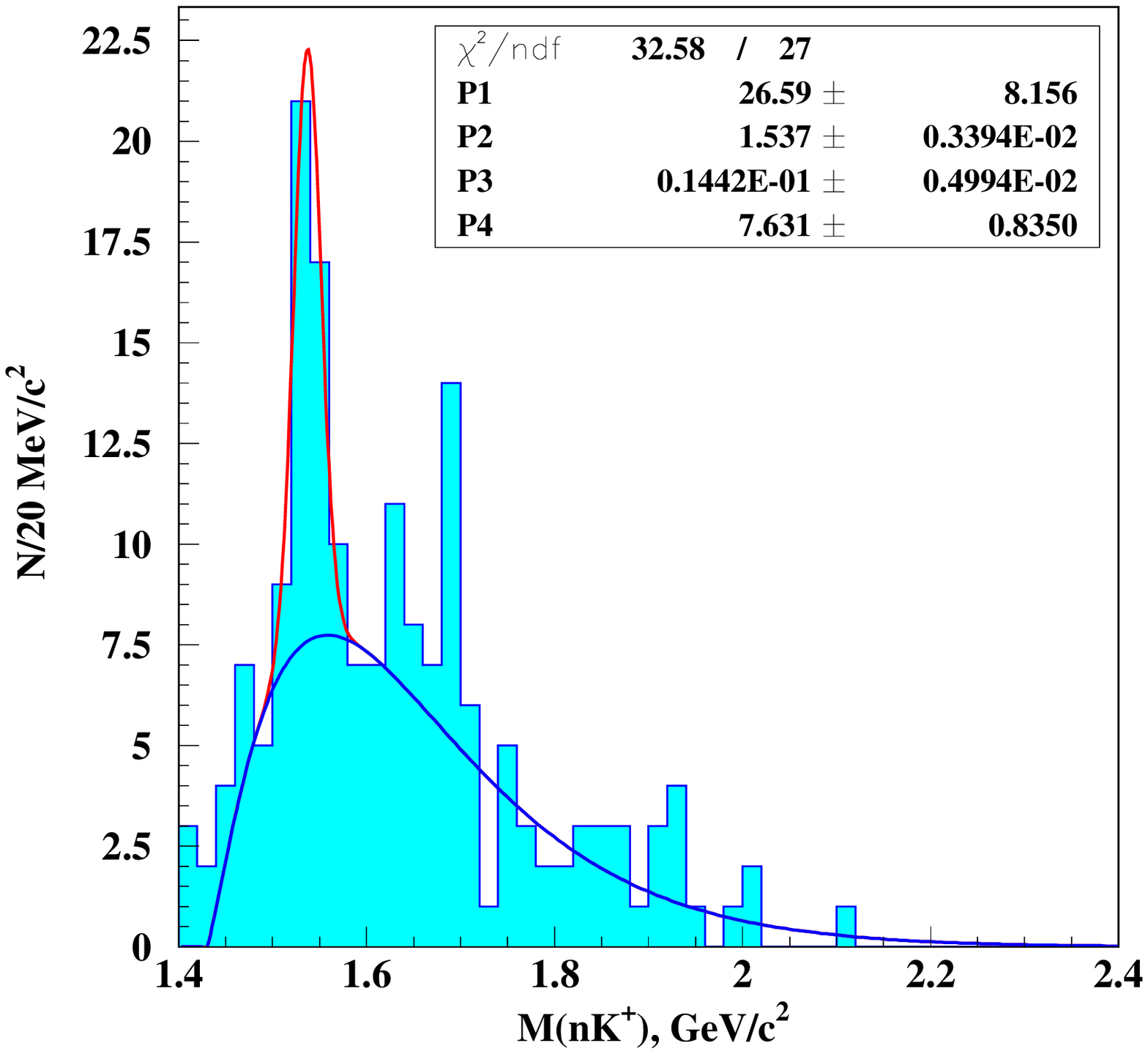}}
\caption{ 
$M_{nK^+}$ invariant mass spectrum  
in the
reaction $\gamma p\rightarrow \pi^+K^-K^+(n)$.
Left panel is for all events. Right panel is for events with
$cos \theta_{cm}>0.5$.
$\theta_{cm}$ 
is the angle between the $\pi^+K^-$  system 
and the photon beam in the center of mass system.
}
\label{theta}
\end{figure}

There are about 40  $\phi$ mesons  in the selected sample.
A  cut on the $M_{K^+K^-}$ invariant mass with
$M_{K^+K^-}>1.040$~GeV/c$^2$ was applied to remove $\phi$ mesons.
The  $M_{nK^+}$ invariant mass spectrum of the remaining  3699 events
is presented in  Fig.\ref{theta}.

To suppress background and extract the signal for 
the $\Theta^+$ photoproduction,
we used the general properties of the production mechanism of $\Theta^+$ 
from a proton target (see Fig.\ref{gamma_p}).
In the tree diagrams  
the $K^-\pi^+$ system moves along the photon beam 
in the center of mass system and $nK^+$ moves in the opposite direction.
In order to select the $t$-channel process illustrated in 
Fig.\ref{gamma_p}, only events with $cos\theta_{cm}>0.5$
were taken for  further analysis.
Since we want to retain both
the resonance $K^{*0}$  production and non-resonance $K^-\pi^+$ continuum
no cuts on $M_{K^-\pi^+}$ were applied.

Fig.\ref{theta} (the right panel) 
presents the  $nK^+$ invariant mass spectrum
of the events with $cos~\theta_{cm}>0.5$. 
This distribution was fitted by a Gaussian function and 
a smooth background.
As suggested by data from this experiment the shape of 
the $nK^+$ invariant mass distribution in the reaction
$\gamma p\rightarrow \pi^+K^-K^+n$  
does not change significantly as a function
of  the $\theta_{cm}$ angle.
For this reason
the shape of the background was obtained from the full data set
of the events (left panel in Fig.\ref{theta}).
The resulting fit yields 27 counts in the peak with the mass
$M=1.54$~GeV/c$^2$ and  width $FWHM=32$ MeV/c$^2$. The mass scale
uncertainty  is estimated as $\pm~10$ MeV/c$^2$. This uncertainty is
mainly due to the energy  calibration of the CLAS detector and 
the electron accelerator.
The statistical significance of this peak is $4.8\pm 0.4~\sigma$
calculated over a window of 80~MeV/c$^2$.
The mass resolution is close to the experimental resolution of CLAS.

As a check, a side band subtraction was carried out
using events from the 
neutron peak ($\pm 3\sigma)$ and background events 
left and right of the neutron peak.
No resonance structures were found in the side band distribution
and the peak parameters of the fit did not 
change after the side band subtraction.

\section{Summary}

Analyses of CLAS photoproduction data firmly establish the existence
of a narrow $S=+1$ exotic baryon  in the $nK^+$ system with a
mass approximately at $1.54$ GeV$/c^2$. 
The statistical significance of the peak
in the invariant mass distribution of the $nK^+$ is $4.8\sigma$ for the
analysis of the reaction $\gamma p\to \pi^+K^-K^+n$, and is
$5.3\sigma$ for the analysis of the reaction $\gamma d\to pK^-K^+n$.
These results are consistent with the $S=+1$ state reported by LEPS
and DIANA collaborations, and with the $5$-quark ($uudd\bar s$)
baryon predicted in the chiral soliton model.

\bibliographystyle{aipproc}   

\end{document}